\documentclass[aps,prl,reprint,superscriptaddress]{revtex4-2}

\begin{document}


\title{\boldmath Comment on ``Search for $\eta'$ Bound Nuclei in the $^{12}\mathrm{C}(\gamma,p)$ Reaction with Simultaneous Detection of Decay Products''}


\author{H.~Fujioka}
	\email{fujioka@phys.titech.ac.jp}
	\affiliation{Department of Physics, Tokyo Institute of Technology, 2-12-1 Ookayama, Meguro, Tokyo 152-8551, Japan}%
\author{K.~Itahashi}%
	\email{itahashi@riken.jp}
	\affiliation{RIKEN Cluster for Pioneering Research, RIKEN, 2-1 Hirosawa, Wako, Saitama 351-0198, Japan}
\author{V.~Metag}%
	\affiliation{II. Physikalisches Institut, Universit\"{a}t Gie\ss en, Heinrich-Buff-Ring 16, 35392 Gie\ss en, Germany}
\author{M.~Nanova}%
	\affiliation{II. Physikalisches Institut, Universit\"{a}t Gie\ss en, Heinrich-Buff-Ring 16, 35392 Gie\ss en, Germany}
\author{Y.~K. Tanaka}
	\affiliation{RIKEN Cluster for Pioneering Research, RIKEN, 2-1 Hirosawa, Wako, Saitama 351-0198, Japan}


\date{\today}


\maketitle

In Ref.~\cite{Tomida2020}, N.~Tomida \textit{et al.}~report
the first result of a missing mass measurement of the $^{12}\mathrm{C}(\gamma,p)$ reactions near the $\eta'$ emission threshold. 
They conducted a simultaneous analysis of the $\eta'$ escape channel and the $\eta'$ absorption by a $^{11}\mathrm{B}$ nucleus followed by emission of an $\eta$ and a proton ($p_s$), making use of the large-acceptance electromagnetic calorimeter, BGOegg. An upper limit was deduced for the formation cross section of an $\eta'$ bound state with subsequent ($\eta + p_s$) decay.
Comparing to a ``theoretical cross section'' the authors of Ref.~\cite{Tomida2020} constrain the real part ($V_0$) of the $\eta'$-$^{11}\mathrm{B}$ optical potential
and the branching fraction ($\text{Br}_{\eta'N\to\eta N}$) in the decay of the $\eta'$-$^{11}\mathrm{B}$ bound state. 
However, we show that their conclusion about $V_0$ and the branching fraction should be reconsidered carefully
in view of unexamined systematic uncertainties in their normalization procedure, as elaborated below.

According to Ref.~\cite{Tomida2020}, the theoretical cross section is defined as: 
\begin{equation}
\left(\frac{d\sigma}{d\Omega} \right)^{\eta+p_s}_{\text{theory}}= 
F ^{(1)}\times \left(\frac{d\sigma}{d\Omega} \right)^{\eta'\text{abs}}_{\text {theory}}\times\mathrm{Br}_{\eta' N\to\eta N}\times P^{\eta p_s}_{srv},
\label{eq1}
\end{equation}
where the ``normalization factor'' $F^{(1)}$ is assumed to be the same as $F^{(2)}$ defined as:
\begin{equation}
F^{(2)}=\left.\left(\frac{d\sigma}{d\Omega} \right)^{\eta'\text{esc}}_{\text{exp}}  \middle/ \left(\frac{d\sigma}{d\Omega} \right)^{\eta'\text{esc}}_{\text{theory}}\right..
\label{eq2}
\end{equation}
Here, for the sake of clarity, we explicitly distinguish the two normalization factors as $F^{(1)}$ and $F^{(2)}$, whereas the common normalization factor $F$ is introduced in Ref.~\cite{Tomida2020}.
While we are not convinced of the validity of the conjecture of $F^{(1)}=F^{(2)}$, 
we hereby raise questions
with regard to the evaluation of $F^{(2)}$.

First, we are concerned with the impact of the imaginary part of the $\eta'$-nucleus potential ($W_0$) on the decomposed cross sections in Eqs.~(\ref{eq1}) and (\ref{eq2}). 
The contributions of the two competing processes, i.e. absorption and escape, in the $\eta'$ unbound region ($E_{\text{ex}}-E_0^{\eta'}>0$) 
are of comparable magnitudes,
reflecting the moderate absorption width of $\eta'$~\cite{Nanova2012, Friedrich2016}, as shown in Fig.~2 of Ref.~\cite{Nagahiro2017}.
The accuracy of $W_0$, which is responsible for the absorption process, is essentially important for the estimation of both $(d\sigma/d\Omega)^{\eta'\text{abs}}_{\text{theory}}$ and $(d\sigma/d\Omega)^{\eta'\text{esc}}_{\text{theory}}$, which appear explicitly in Eqs.~(\ref{eq1}) and (\ref{eq2}).
While $W_0$ was fixed at $-12\,\mathrm{MeV}$ in accordance with the CBELSA/TAPS result ($-13\pm 3\pm 3\,\mathrm{MeV}$)~\cite{Friedrich2016}, 
adding the statistical and systematic errors in quadrature leads to a range of $|W_0|$ values between 9 and $17\,\mathrm{MeV}$, which has to be taken into account in evaluating the uncertainty of the l.h.s.~of Eq.~(\ref{eq1}), owing to systematic uncertainties of $F^{(2)}$ and $(d\sigma/d\Omega)^{\eta'\text{abs}}_{\text{theory}}$~\footnote{A decrease in the magnitude of $W_0$ will decrease
 $(d\sigma/d\Omega)^{\eta'\text{abs}}_{\text{theory}}/(d\sigma/d\Omega)^{\eta'\text{esc}}_{\text{theory}}$. 
In the Reply to our Comment by N.~Tomida~\textit{et al.} (unpublished), the authors state ``the lower fluctuation of $(d\sigma/d\Omega)^{\eta'\text{abs}}_{\text{theory}}/(d\sigma/d\Omega)^{\eta'\text{esc}}_{\text{theory}}$, which influences the upper limit of $\text{Br}_{\eta' N\to \eta N}$, is estimated to be 16\% and 19\% in the case of $V_0=-20$ and $-100\,\mathrm{MeV}$, respectively, within the statistical uncertainty of $W_0$ in Ref.~\cite{Friedrich2016}''.
Although the authors conclude ``these values are smaller than the statistical uncertainty of $(d\sigma/d\Omega)^{\eta'\text{esc}}_{\text{exp}}$ of 26\%
shown with hatch in Fig.~4 of Ref.~\cite{Tomida2020}'', 
we stress the fluctuation of 16--19\%, being comparable with the statistical uncertainty of 26\%, should be treated as a systematic uncertainty in the determination of the branching ratio}.

Second, we would like to point out a wide range of the momentum transfer,
due to the polar angle ($\theta_p$) coverage of the ejectile protons as well as the range of the incident photon energy ($E_\gamma$) between 1.3 and $2.4\,\mathrm{GeV}$. 
For instance, the momentum transfer at the threshold energy ($E_{\text{ex}}-E_0^{\eta'}=0$) is $0.47\mbox{--}0.49\,\mathrm{GeV}/c$ at $E_\gamma=1.3\,\mathrm{GeV}$, and $0.24\mbox{--}0.36\,\mathrm{GeV}/c$ at $E_\gamma=2.4\,\mathrm{GeV}$, for the $\theta_p$ coverage of $0.9^\circ\mbox{--}6.8^\circ$.
Note that the strengths of different angular momentum components of the $\eta'$-nucleus formation cross section behave differently as a function of the momentum transfer 
(see, e.g., Fig.~1 of Ref.~\cite{Nagahiro2017}, exhibiting scattering-angle dependence of the formation cross section).
Therefore, it requires justification to evaluate $F^{(2)}$ as the ratio of the averaged cross sections instead of Eq.~(\ref{eq2}).
It would be more appropriate to evaluate $F^{(2)}$ as a function of $E_\gamma$ and $\theta_p$.
Indeed,  Fig.~3 in Ref.~\cite{Tomida2020} implies a possible $E_\gamma$ dependence of $F^{(2)}$.

In summary, the ``suppressed normalization ambiguity'', stated in the abstract of Ref.~\cite{Tomida2020}, may stem from 
these simplifications, and, consequently, $V_0$ and $\text{Br}_{\eta'N\to\eta N}$ may be overconstrained.
We emphasize that the two aforementioned aspects are much less significant in the inclusive measurement
of the $^{12}\mathrm{C}(p,d)$ reaction by the $\eta$-PRiME/Super-FRS Collaboration~\cite{PRL2016,PRC2018}
with a fixed beam energy of $2.5\,\mathrm{GeV}$, resulting in 
a narrow range of the momentum transfer ($0.48\mbox{--}0.49\,\mathrm{GeV}/c$) at the threshold energy.
While the authors of Refs.~\cite{PRL2016,PRC2018} have not evaluated $F$ in a similar way, 
the $\mu_{95}$ parameter defined in Refs.~\cite{PRL2016,PRC2018} is nothing but the normalization factor,
which must be differentiated from the $F^{(2)}$ factor in Ref.~\cite{Tomida2020}, obtained for the different kinematical conditions including the reaction itself.
The boundaries of the allowed $(V_0,W_0)$ regions for various $\mu_{95}$ parameters are explicitly depicted in Fig.~4 of Ref.~\cite{PRL2016} and Fig.~11 of Ref.~\cite{PRC2018}.

\bibliography{comment_BGOegg}
\end{document}